\documentclass[10pt, conference]{IEEEtran}

\usepackage{amssymb}
\usepackage{amsmath}

\usepackage[colorlinks = true,
  linkcolor = blue,
  citecolor = blue,
  urlcolor = blue]{hyperref}

\usepackage{graphicx}
\usepackage{booktabs}
\usepackage{multirow}
\usepackage{tabularx}
\usepackage{enumitem}
\usepackage{threeparttable}
\usepackage{balance}
\usepackage{url}

\usepackage{fancyhdr} 
\fancypagestyle{firststyle}
{
\fancyhf{}
\fancyfoot[C]{\scriptsize{Proceedings of the IEEE 33rd International Requirements Engineering Conference Workshops (REW 2025), Valencia, IEEE, 2025, pp.~215--222. \\ This version is the authors' copy. The publisher's definite version is available online via \url{https://doi.org/10.1109/REW66121.2025.00034}.}}
}

\begin{document}

\title{A Mapping Analysis of Requirements \\ Between the CRA and the GDPR}

\author{
\IEEEauthorblockN{Jukka Ruohonen}
\IEEEauthorblockA{University of Southern Denmark \\
Email: juk@mmmi.sdu.dk}
\and
\IEEEauthorblockN{Kalle Hjerppe}
\IEEEauthorblockA{University of Turku \\
Email: kphjer@utu.fi}
\and
\IEEEauthorblockN{Eun-Young Kang}
\IEEEauthorblockA{University of Southern Denmark \\
Email: eyk@mmmi.sdu.dk}
}

\maketitle

\begin{abstract}
A new Cyber Resilience Act (CRA) was recently agreed upon in the European Union
(EU). The paper examines and elaborates what new requirements the CRA entails by
contrasting it with the older General Data Protection Regulation
(GDPR). According to the results, there are overlaps in terms confidentiality,
integrity, and availability guarantees, data minimization, traceability, data
erasure, and security testing. The CRA's seven new essential requirements
originate from obligations to (1) ship products without known exploitable
vulnerabilities and (2) with secure defaults, to (3) provide security patches
typically for a minimum of five years, to (4) minimize attack surfaces, to (5)
develop and enable exploitation mitigation techniques, to (6) establish a
software bill of materials (SBOM), and to (7) improve vulnerability
coordination, including a mandate to establish a coordinated vulnerability
disclosure policy. With these results and an accompanying discussion, the paper
contributes to requirements engineering research specialized into legal
requirements, demonstrating how new laws may affect existing requirements.
\end{abstract}


\section{Introduction}

\thispagestyle{firststyle} 

The paper examines the legal requirements imposed by the EU's CRA and GDPR
regulations.\footnote{~Regulation (EU) 2024/2847 and Regulation (EU) 2016/679,
respectively.} The former was agreed upon in December 2024, while the latter
came enforceable in 2018. As two regulations are considered, a natural
motivation is to ask (Q.1): what is common between them? This motivation aligns
well with requirements engineering because overlapping requirements, whether due
to a repetition by different stakeholders or something else, is a classical
topic in requirements engineering~\cite{Ellis17, Nguyen13}. Among other things,
overlapping requirements may cause redundancy; therefore, the motivation can be
sharpened by considering an imaginary software or system that is already
compliant with the GDPR's legal requirements. Then, after it is known what is
common, it can be deduced what is new. In other words (Q.2): what new legal
cyber security requirements the CRA brings?

Redundancy is a classical concept also in legal scholarship and related
fields~\cite{Golden16, Landau69}. The reason is understandable: laws often
repeat themselves. This legal repetition is partially explained by lawmaking,
which is often an iterative endeavour in a sense that new laws often (but not
always) build upon existing laws. The concept of an architecture provides a
related analogy; the ``high-dimensionality of the architecture of both law and
computer science implies that choices made at any point of the system will
ripple through the entire system, resulting in bugs or new
features''~\cite[p.~2]{Hildebrandt20}. Given this insightful quotation, the
focus is on new requirements but not on new features; the scope is restricted to
requirements engineering alone, without delving into a question on how these
might be designed and implemented, or, given the earlier remark about some
existing GDPR-compliant implementation, how it should be changed. Nevertheless,
the quotation's point about ripple effects provides another motivation for the
paper. As was envisioned in an early requirements engineering research regarding
the GDPR~\cite{Hjerppe19RE}, and as has also generally been observed
~\cite{Massey19, Otto07}, also legal requirements change over time, whether due
to changing interpretations and guidelines by regulators, court practice and
case law, enactment of new laws or deprecation of existing laws, or something
else.

\begin{figure}[th!b]
\centering
\includegraphics[width=\linewidth, height=1.5cm]{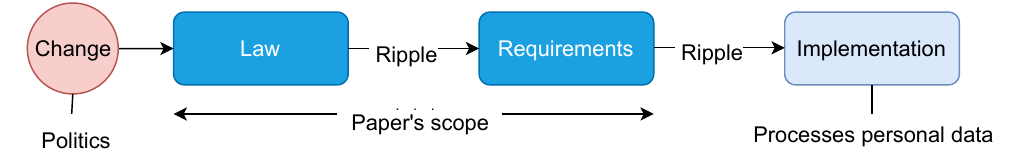}
\caption{Changes, Ripples, and the Paper's Scope}
\label{fig: change}
\end{figure}

Overlaps, redundancy, duplication, and related terms justify also what is meant
by a mapping analysis in the present context; the focus is on the intersection
of the legal requirements imposed by the two EU regulations on one hand, and the
relative complement of the requirements on the other hand. The latter refers to
the CRA's new requirements that are not also a part of the GDPR. By using this
set theory terminology, a necessary framing follows: (a)~the focus is on
implementations that process personal data~(see Fig.~\ref{fig: change}). In this
regard, it should be emphasized that the CRA applies also to non-personal
data. This framing may also explain the legal redundancy; strict references to
the GDPR are not possible in most cases because the CRA's scope is much wider in
terms of cyber security. A~further framing is important to mention: (b) the CRA
is only considered in terms of its so-called essential cyber security
requirements---a term elaborated later in Subsection~\ref{subsec: conformance
  and compliance}. Furthermore, (c) the GDPR is only considered in terms of its
intersection with the CRA. In light of the existing and related work briefly
outlined in the opening Section~\ref{sec: related work}, this choice is
justifiable due to the vast amount of research on the GDPR. There is no reason
to commit to redundancy also in science.

\section{Related Work}\label{sec: related work}

The GDPR has received a considerable attention after its enactment, including in
requirements engineering and associated computing disciplines~\cite{Birrell24,
  Hjerppe19RE, NegriRibalta24, Saltarella23}. Due to the CRA and the mapping
approach, the related work also expands toward cyber security research. In this
regard, it can be mentioned that cyber security violations rank high also with
respect to GDPR enforcement fines, although still lower than the regulation's
basic legal principles~\cite{Ruohonen22IS}. Such cyber security violations
likely also continue today and in the nearby future because (personal) data
breaches and ransomware operations are currently the two most pressing
high-impact risks for many organizations~\cite{ENISA24a}. These also illustrate
the basic concepts discussed later in Subsection~\ref{subsec: cia triad}: at
minimum, the former typically violate confidentiality and the latter
availability.

Due to other recent ripples in the EU's legal framework for cyber security, the
GDPR has also been considered in conjunction with other recent cyber security
laws~\cite{Mortensen24}. Thus far, however, existing research on the CRA is
scant but not entirely absent, the likely explanation being simple; at the time
of writing, the regulation is new and the details are still being worked on in
the member states. Regarding research---and in addition to legal overviews
\cite{Mueck25}, the CRA's new obligations for vulnerability coordination and
disclosure have been discussed~\cite{Ruohonen24IFIPSEC}. Recently, the alignment
of the CRA's essential requirements and the MITRE's
ATT\&CK\textsuperscript{\scriptsize\textregistered} mitigations has been
evaluated~\cite{Ruohonen25ESPRE}. There is also work on the CRA's still unclear
implications upon open source software~\cite{Colonna25}. The regulation's
requirement about secure default configurations has also been implicitly
addressed through a literature review~\cite{Ruohonen25JISA}. These and
potentially other outlying examples notwithstanding, the volume of existing
research is still low. For this reason, the CRA also takes a priority in the
mapping in the form of Q.2. Further related work can be pointed out by
considering some of the core concepts together with additional framings.

\section{Concepts and Further Framings}

\subsection{Cyber Security and Data Protection}\label{subsec: cyber security and data protection}

Cyber security and data protection have an intrinsic relationship already
because confidentiality belongs to the former. Indeed, the GDPR contains also
explicit obligations to protect personal data of natural persons from cyber
security threats. When considering the intersection of these two regulations,
the framing sharpens further toward the GDPR's cyber security obligations; the
regulation's many other requirements are framed out. While cyber security is
fully in the CRA's scope, the framing toward existing but imaginary
implementations further restricts the mapping toward technical aspects, which
are also what the essential requirements are about. Because the CRA is a
product-specific regulation, whereas the GDPR applies to anything processing
personal data, including organizations, the framing is toward products that
process personal data. Such products must comply with both regulations.

As for cyber security, it can be noted that the concept is much broader than
older related concepts, such as information security within which information
(including personal data) is the asset to be protected. In both regulations, but
particularly in the CRA, the broader scope is seen in the use of the resilience
concept, among other things. As will be seen, there are also new mandates that
consider the availability of other networks and services as an asset that should
not be harmed by products.

\subsection{Requirement Types}

There is a classical distinction between functional and non-functional
requirements in software engineering. The former refer generally to something
that a software should do, whereas non-functional requirements are more general
quality attributes of the software, which also constraint the
software~\cite{IEEE04}. Both security and privacy have usually been categorized
as non-functional requirements~\cite{Horkoff19, Khatter13, Mairiza10}. The same
applies to data protection.  While some have seen the GDPR as a series of
non-functional requirements~\cite{Li19}, others have perceived it to contain
both functional and non-functional requirements~\cite{Hjerppe19RE,
  NegriRibalta24}. Similar ambiguities apply in terms of cyber security. For
instance, once broken down to smaller details, encryption can be seen to contain
both functional and non-functional requirements~\cite{Farkhani06}. By following
this line of reasoning, the mapping analysis approaches the distinction from a
perspective of composition; the overall non-functional requirement of cyber
security can be seen to contain various functional requirements describing what
an implementation should do. These functional requirements are understood as
being representable in a flow diagram operating at the same abstraction level as
a given non-functional requirement, which, in contrast, is not a process. With
respect to functional requirements, a further separation is done between
functional requirements for software, systems, and other implementations and
``organizational'' requirements with clear functions. As soon discussed in
Subsection~\ref{subsec: risks}, the distinction is relevant because the CRA's
functional requirements are dependent on the organizational (functional)
requirement of risk analysis.

A further clarification is required: the paper operates with legal requirements,
which are generally about something that a law requires for software, systems,
and other implementations. The term legal requirement is important for three
reasons. First, it separates law-imposed requirements from ``conventional''
requirements, such as those elicited from clients and customers. With respect to
non-functional requirements in general, legal requirements are particularly
something that are non-negotiable~\cite{vanVliet07}. Second, compliance with
legal requirements can be seen as an overall non-functional requirement from
which, if possible, functional requirements should be drawn. As non-functional
requirements too should be verifiable~\cite{vanVliet07}, this point reiterates
the point about processes. Third, legal requirements are also a distinct genre
within requirements engineering research. Despite various models, frameworks,
techniques, and tools, the challenges in this genre are
well-recognized~\text{\cite{Hjerppe19RE, Otto07}}. In particular, legal
requirements are methodologically challenging in requirements engineering
research because these require an interdisciplinary approach. Although different
ontologies and related techniques have been presented~\cite{Tamburri20},
elicitation of legal requirements necessitates actually reading and interpreting
law, no more or no less. This truism explains also why a collaboration between
engineers and lawyers is often desirable in the industry when dealing with legal
requirements~\cite{Hjerppe19RE}. Reading and interpreting are also what the
paper's methodological approach for the mapping analysis is about---as is the
``methodology'' in legal scholarship in general. While these provide a basic
toolbox to make the mappings, a step toward more traditional requirements
engineering is taken by tentatively further interpreting how the legal
requirements identified translate into functional requirements for software,
systems, and other implementations

\subsection{Conformance and Compliance}\label{subsec: conformance and compliance}

The CRA categorized products into three groups: ``normal'' products (Article 6),
important products (Article 7), and critical products (Article 8). The
``normal'' ones must comply with the essential cyber security regulations
enumerated in the regulation's Annex I. These are also the paper's focus. Then,
important products, which are listed in Annex III, must further comply with
conformity assessment procedures specified in Article 32. While there are
different options in these procedures, a typical choice for many vendors will
likely be a self-assessment of conformity, as has also been common with the EU's
product safety laws~\cite{Ruohonen22ICLR}. It is also possible to choose a
stricter procedure involving third-party audits. Such audits are required from
the critical products. These are linked to the cyber security certification
scheme established in the Cybersecurity Act, that is, Regulation (EU)
2019/881. The audits themselves will be supported by standards~\cite{ENISAJRC24,
  Mueck25}. For the paper's purposes, it is important to emphasize that, with a
few exceptions, practically all information technology products need to comply
with the CRA's essential requirements.

\subsection{Requirements and Risks}\label{subsec: risks}

Both the CRA and the GDPR are risk-based regulations, meaning that measures
taken, including requirements elicited from the regulations, should be balanced
against results from risk analyses. Given the exhaustive body of knowledge on
cyber security risk management~\cite{Talbot09}, which aligns with the CRA's
obligations, it suffices to remark that the GDPR's risk analysis mandates, which
include privacy impact assessments, can be done analogously by considering risks
to natural persons specifically. In fact, classical risk matrices involving
impacts and probabilities have been recommended in the GDPR
context~\cite{ENISA17}. This point is important to emphasize because risks are a
classical way to prioritize security requirements, including countermeasures and
security design solutions in general~\cite{Mead05}. Although such a
prioritization is left for further work as it requires an existing
implementation or at least a design for it, it should be understood that not all
of the legal requirements carry the same weight in each and every case.

\section{Overlaps}

In what follows, the overlapping legal requirements are discussed and
elaborated. The overlaps should be understood to apply with respect to the
framings noted in the introduction. Before continuing, therefore, it should be
understood that the CRA applies to all product with a ``a direct or indirect
logical or physical data connection to a device or network'', as stated in
Article~2(1). Thus, anything with a network connection is in the scope. The
notable exclusions are maritime equipment and ships, medical devices, motor
vehicles, and cloud computing, including software, platform, and infrastructure
as a service. The likely reason is that these excluded domains are covered in
their own specific laws. With the framing toward the GDPR, a context of the
overlaps could thus range from Internet of things~(IoT) devices to information
systems and beyond.

\subsection{The CIA Triad}\label{subsec: cia triad}

\begin{figure}[th!b]
\centering
\includegraphics[width=\linewidth, height=9cm]{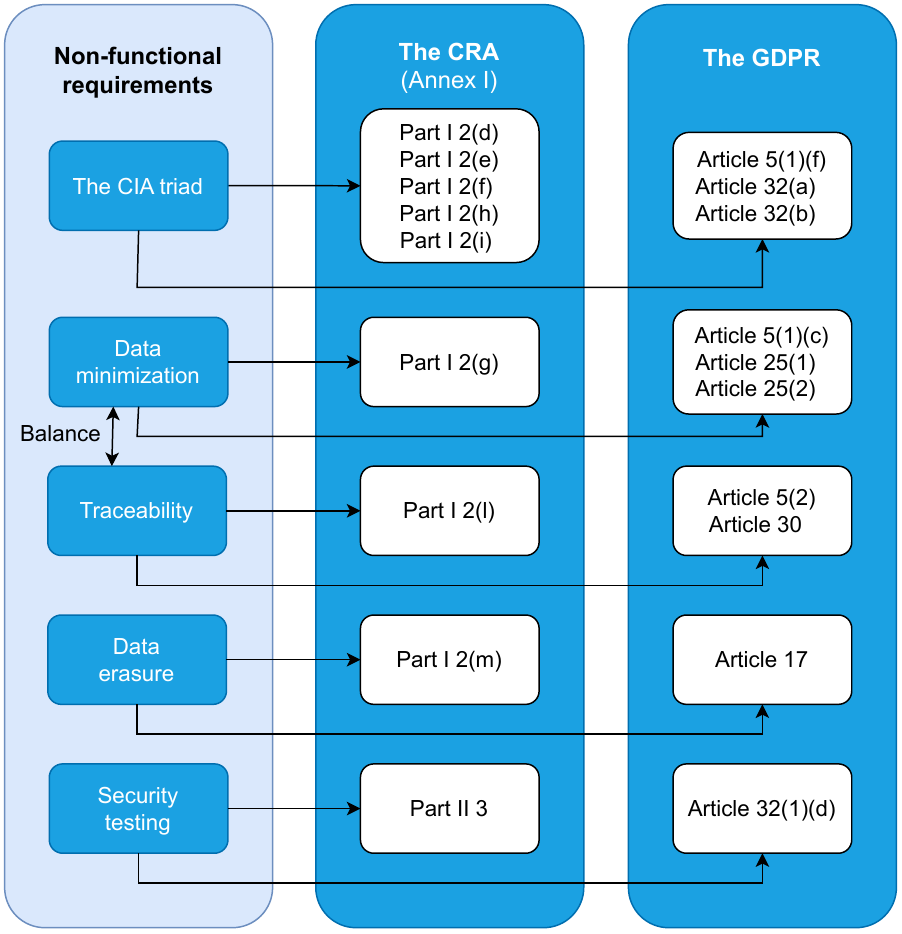}
\caption{Overlaps Between the CRA's and the GDPR's Requirements}
\label{fig: overlaps}
\end{figure}

The classical confidentiality, integrity, and availability (CIA) triad appears
in both regulations (see~Fig.~\ref{fig: overlaps}). Like cyber security
itself~\cite{Khatter13}, the CIA triad has usually been interpreted as a
non-functional requirement~\cite{Mairiza10}. This interpretation seems suitable
also for the present work because neither of the two regulations define any
specific technical functionality on how to guarantee confidentiality, integrity,
and availability, the exception of encryption perhaps~notwithstanding.

With respect to the GDPR, the CIA triad appears literally in paragraphs (b) and
(c) in Article 32. In addition to mentioning the triad explicitly, the
regulation also obliges to ensure resilience, which includes an ability to
restore availability and access to personal data during incident management. The
article's paragraph (a) further mentions encryption and pseudonymization, which
are both related to confidentiality. As the latter concept is specific to the
GDPR, it is omitted from the mapping analysis without a notable loss of
specificity.

Then, regarding the CRA, the CIA triad is addressed through four essential
requirements. Of these, the confidentiality and integrity requirements use a
wording about ``stored, transmitted or otherwise processed''. With respect to
data, whether personal or non-personal, both of which are in the CRA's (but not
in the GDPR's) scope, the wording reflects the commonplace security taxonomy of
``data at rest'', ``data in transmission'', and ``data in
processing''~\cite{Eltaeib21, Jakobik16}. Encryption is a typical security
requirement for the first two concepts.

With respect to confidentiality, encryption is again explicitly mentioned in the
CRA analogously to the GDPR. However, regarding integrity, the scope is wider
because the integrity guarantees should cover also ``commands, programs and
configuration'' against manipulation or modification by unauthorized parties. Of
these three concepts, unauthorized commands could be reflected against
structured query language (SQL) injections or even buffer overflows, while
unauthorized programs might cover an execution of unauthorized third-party code,
as has been common in the web domain~\cite{Ruohonen18IFIPSEC}, among other
domains. Furthermore, there is a specific essential requirement about ensuring a
protection against unauthorized access, which too is related to confidentiality
and integrity. Authentication and identity management are mentioned as examples.

Availability too should be guaranteed, as always, and again resilience is
mentioned also in the CRA as a concept to address post-incident recovery
solutions together with mitigative measures against denial-of-service (DoS)
attacks. In addition, there is an interesting and relevant new requirement, as
specified in paragraph 2(i) in Part I of Annex I, to minimize any negative
effects upon products and connected devices themselves ``on the availability of
services provided by other devices or networks''. This essential requirement can
be reflected against a concept of cascading risks, failures, and incidents; a
single event may lead to further events, whether in terms of cyber security or
more generally~\text{\cite{ENISA18, Little02}}. A prime historical example would
be the Mirai botnet that was built upon compromised IoT
devices~\cite{Margolis17}. It remains to be seen whether the CRA may decrease a
probability of such botnets emerging in the future, and whether a producer or
producers could be held liable for enabling such emergencies through gross
negligence of the essential cyber security requirements.

\subsection{Data Minimization}

Data minimization is an important principle and an obligation in the GDPR; one
should only collect a minimal amount of data necessary to fulfill a given
purpose. Interestingly enough, the CRA specifies the same obligation; only what
is necessary to a purpose should be collected. The mandate is interesting
because the CRA applies also to non-personal data. In any case, data
minimization should be balanced with traceability.

\subsection{Traceability}

Both regulations impose a traceability requirement. Although logging is the
obvious solution to fulfill the requirement~\cite{Hjerppe19RE}, the term
traceability in itself is slightly better because the GDPR applies also to
manual processing of personal data. In any case, the CRA's paragraph 2(l) in
Part~I of Annex I requires ``recording and monitoring relevant internal
activity, including the access to or modification of data, services or
functions''. Thus, not only should the GDPR-specific operations, such as an
erasure or a rectification of personal data, be logged but the logging should in
principle apply to all CRUD operations, to use the common abbreviation for
create, read, update, and delete. As said, however, balancing should be likely
still done in many cases, and also for scalability reasons~\cite{Rong17}. A
further point is that the wording used about an access to services and functions
indicates also other considerations. To use the IoT domain again as an example:
a device's connection to a network might be logged.

\begin{figure}[th!b]
\centering
\includegraphics[width=7cm, height=9.0cm]{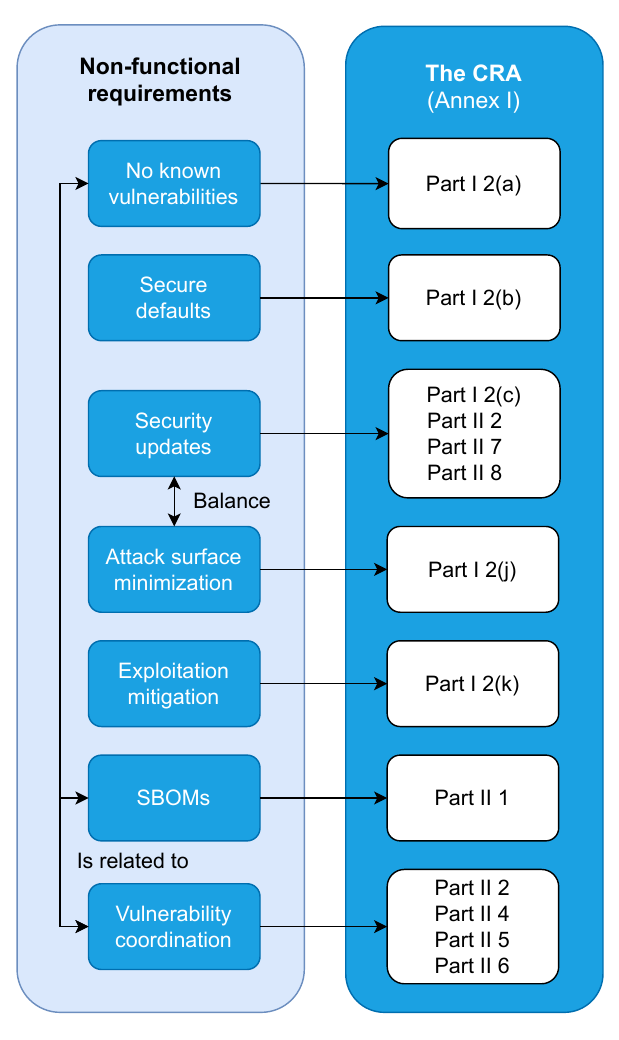}
\caption{The CRA's New Essential Requirements}
\label{fig: cra new}
\end{figure}

\subsection{Data Erasure}\label{subsec: data erasure}

Although the right to personal data protection is not absolute, the GDPR grants
a right to delete one's personal data in many cases. Also the CRA contains this
requirement but with a more general wording about a secure and easy removal ``on
a permanent basis all data and settings''. Although settings should be deleted
too, the term data erasure is used in Fig.~\ref{fig: overlaps} due to the
framing. The emphasis on permanent and secure deletion reiterates a classical
topic in cyber security~\cite{Reardon13}. The erasure requirement is also
generally related to the CRA's mandate to support products for a minimum of five
years in most cases. The relation between life cycles and the erasure
requirement comes from a broader contemplation about sustainability, the green
transition, and second-hand markets.

\subsection{Security Testing}

\begin{figure*}[t!]
\centering
\includegraphics[width=\linewidth, height=9.7cm]{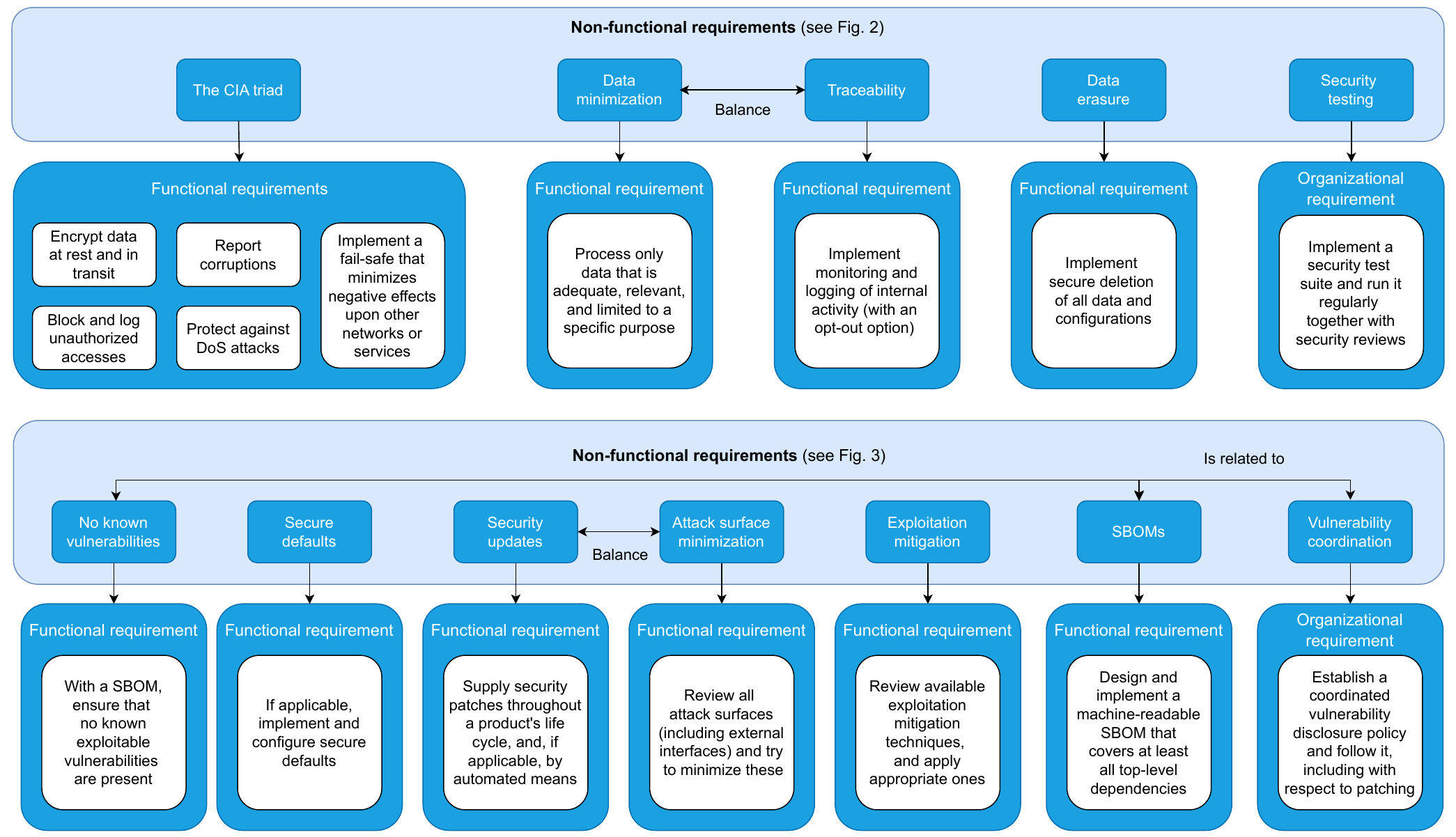}
\caption{Non-Functional, Functional, and Organizational Requirements Identified
  from the CRA's Essential Cyber Security Requirements}
\label{fig: functional}
\end{figure*}

Both the CRA and the GDPR require security testing. In terms of the latter, the
requirement is specified in Article~32(d), which states that there should be ``a
process for regularly testing, assessing and evaluating the effectiveness of
technical and organisational measures for ensuring the security''. The coverage
of organizational security reflects the conceptual issues that were noted in
Subsection~\ref{subsec: cyber security and data protection}. In the CRA a
wording ``effective and regular tests and reviews'' is used. In other words,
also security reviews should be done alongside~testing.

\section{The CRA's New Essential Requirements}

The new cyber security requirements imposed by the CRA regulation can be grouped
into seven groups. These are summarized in Fig.~\ref{fig: cra new} and briefly
elaborated in what follows. To summarize the points raised, Fig.~\ref{fig:
  functional} additionally displays the functional requirements explicitly
identifiable from the CRA. While risk analysis is omitted from the figure for
brevity, it too is a functional and organizational requirement. In addition,
many of the CRA's essential requirements, including the earlier ones in
Fig.~\ref{fig: overlaps}, must be documented according to Article~13.

\subsection{No Known Vulnerabilities}

The first new essential requirement is simple enough: all products must ``be
made available on the market without known exploitable vulnerabilities''. The
wording about exploitability gives some leeway for vendors, but otherwise the
legal mandate is on the side of straightforward functional requirements. To be
able to fulfill the requirement, a vendor obviously must know which components
it uses, whether software or hardware components. In terms of software, the
requirement therefore aligns with a mandate to establish a software bill of
materials for each product. There is also a philosophical angle to the
requirement: known by whom? While the regulation's future guidelines and its
enforcement may clarify the question, presently it would seem sufficient to
focus on the major international vulnerability databases.

\subsection{Secure Defaults}

The CRA mandates that all new products entering the EU's internal market should
be shipped with secure default configurations. Though, an opt-option is given
with a wording ``unless otherwise agreed between manufacturer and business user
in relation to a tailor-made product''. In any case, as this requirement has
been discussed in detail recently~\cite{Ruohonen25JISA}, it suffices to remark
that the requirement is a classical security design principle. For instance, the
OpenBSD operating system has long endorsed the
principle~\cite{OpenBSD25}. Having said that, the IoT domain's track record
exemplifies that the principle has often been also ignored or even
abused~\cite{Cavalli24}. Therefore, the CRA's new essential requirement can be
generally welcomed.

\subsection{Security Updates}\label{subsec: security updates}


All products should be generally supported for a minimum of five years according
to the CRA. Though, the wordings in recital 60 and Article 13(8) are vague,
suggesting that a shorter support period is also possible in some
cases. Regardless, the essential requirement is that security patches for
vulnerabilities should be provided free of charge throughout a product's life
cycle. Furthermore, vulnerabilities should be addressed ``where applicable,
through automatic security updates''. Even when applicable, the automated
updating functionality should come with ``a clear and easy-to-use opt-out
mechanism'' as well as notifications to users about updates and a further option
to ``temporarily postpone them''. Regardless whether automation is used or not,
updates should be securely distributed. Furthermore, ``where technically
feasible, new security updates shall be provided separately from functionality
updates''. These clarifications and options are summarized in Fig.~\ref{fig:
  updates}.

Even though optional, automatic updates can be seen as particularly relevant for
improving the IoT domain's cyber security posture~\cite{Cavalli24}. The
provisions and options provided also indicate that the CRA seems to have taken
into account \text{most---or} even all---of the deadly sins associated with
software updates, as elaborated in the literature~\cite{Howard10}. However, the
CRA's wide scope, which includes everything from smartphones to industrial
control systems, will likely bring new challenges for some vendors. In addition,
there are always risks involved too in updating; in fact, annual surveys
indicate that faulty updates have been relatively common in some industry
sectors~\cite{ENISA24b}. Faulty updates have also led to some memorable, severe,
but non-security incidents over the years~\cite{Nolan24}. Furthermore,
regressions are unfortunately common in many software projects, including
critical ones, such as the Linux kernel~\cite{Ruohonen24EASEa}. These and other
potential problems warrant also further research in software engineering and
beyond.

\begin{figure}[th!b]
\centering
\includegraphics[width=\linewidth, height=3.2cm]{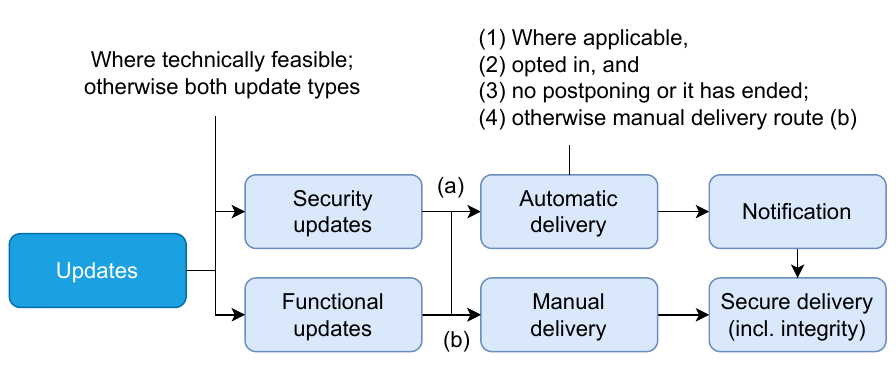}
\caption{Updating Requirements}
\label{fig: updates}
\end{figure}

\subsection{Attack Surface Minimization}

There is also an essential and explicit requirement to minimize attack
surfaces. The CRA mentions external interfaces in this regard, but nothing
beyond that is available. Although there is plenty of research in this area,
including in the IoT context~\cite{Sotiropoulos23}, a systematization of
knowledge might be beneficial now that the topic is covered also in a new
law. It can be also noted that this requirement seems to contradict slightly the
requirement of security updates. In other words, the functionality in
Fig.~\ref{fig: updates} also enlarges attack surfaces.

\subsection{Exploitation Mitigation}

The CRA requires that appropriate exploitation mitigation techniques should be
designed, developed, and enabled. This essential requirement aligns partially
with a security design principle of turning security features
on~\cite{Ruohonen25JISA}. However, the CRA does not say anything specific about
the techniques.

\subsection{SBOMs}

As already noted, SBOMs must be implemented for all products. While useful also
for vendors and their supply-chain management, they may be used for enforcement
purposes too, given that a vendor is obliged to deliver its SBOMs to supervising
authorities upon request according to Article~13(25). The Article's context and
motivation are about potential large-scale sweeps of open source software
dependencies. Confidentiality should be guaranteed when dealing with authorities
and vendors are not obliged to make their SBOMs public.

\subsection{Vulnerability Coordination}

Finally, the CRA brings some new requirements for vulnerability coordination,
including vulnerability disclosure. As this requirement has already been
considered in existing work~\cite{Ruohonen24IFIPSEC}, it suffices to merely
enumerate some of the particular requirements: vulnerabilities should be fixed
without delay, updates fixing them should be disseminated without delay, they
should be documented in conjunction with SBOMs, security advisories should be
delivered to users alongside public disclosures, information sharing about
vulnerabilities should be facilitated, coordinated vulnerability disclosure
policies should be established and followed, and so forth and so on.

\section{Discussion}

\subsection{Conclusion}

The paper examined the new CRA regulation by comparing and contrasting it with
the older GDPR. While the CRA's scope is much wider, including most products
with a networking functionality and covering also processing of non-personal
data, the mapping study conducted revealed also similarities. Notably, the CIA
triad guarantees, data minimization, traceability, data erasure, and security
testing appear in both regulations, although a verbatim equivalence is not
present (Q.1). Regarding non-overlapping requirements, the CRA imposes mandates
to launch products without known vulnerabilities and with secure default
configurations, to provide security updates, to minimize attack surfaces, to
enable appropriate exploitation mitigation techniques, to implement SBOMs, and
to follow established vulnerability coordination practices (Q.2). All in all,
the requirements can be seen as sensible for the regulation's goal of improving
cyber security in Europe. Nor do these seem overly burdensome to design and
implement, although the CRA's broad scope may cause some problems in some
industry sectors with specialized products.

\subsection{Further Overlaps}

As has been pointed out~\cite{Mueck25}, there is a further overlap between the
two regulations and the radio equipment Directive 2014/53/EU. There are three
obligations specified in the directive's Article 3(3)(d), (e), and (f) that
overlap with the CRA and the GDPR: personal data should be protected, fraud
prevention should be implemented when possible, and an ``equipment does not harm
the network or its functioning nor misuse network resources, thereby causing an
unacceptable degradation of service''. The obligation quoted resembles the point about cascading effects noted in Subsection~\ref{subsec: cia triad}.

As has again been pointed out~\cite{Ruohonen24PRR}, the GDPR also overlaps
slightly with the EU's recent Directive (EU) 2022/2555, which is commonly known
as the NIS2 directive. The overlap comes from incident notifications about
serious personal data breaches to public authorities. Although a direct overlap
is not present, also the CRA imposes incident notification obligations. The
regulation's Article~14 mandates manufacturers to report about actively
exploited vulnerabilities as well as severe incidents to the supervising
authorities. While the deadlines of reporting align with those specified in the
NIS2 directive~\cite{Ruohonen25COSE}, the definitions for severe incidents are
different. According to the CRA's Article~14(5), these refer to those that
affect the CIA guarantees of products on one hand, and those ``leading to the
introduction or execution of malicious code in a product with digital elements
or in the network and information systems of a user of the product'' on the
other hand. Once again, the theme of cascading effects is present in this direct
quotation. As the NIS2 directive obliges operators of critical infrastructures
to notify about other incident types too and conduct comprehensive risk
analyses, there is also a more broad overlap with the CRA. Seeking synergies
between these two cyber security laws may be worthwhile in practice.

Furthermore, there is an overlap with Regulation (EU) 2023/1670 addressing
mobile phones, including smartphones. This overlap is about product life cycles,
including the obligation to provide security updates discussed in
Subsection~\ref{subsec: security updates} and the sustainability theme noted in
Subsection~\ref{subsec: data erasure}. That is, the mobile phone regulation
mandates phone manufacturers to provide operating system updates for a minimum
of five years.

\subsection{Further Work}

With the results from the mapping analysis, the paper contributes to the early
efforts to build a reusable body of knowledge on legal
requirements~\cite{Otto07}. Further work is required to go beyond requirements;
to evaluate what must be done or changed in the domain of actual software,
systems, and other implementations. However, due to the CRA's broad scope,
one-size-fits-all solutions or recommendations are presumably difficult to
postulate. Thus, case studies would offer a good path forward to better
contextualize the requirements elicited from the regulation. Another option
would be to maintain a high level of abstraction and evaluate how the CRA's
requirements align with existing cyber security design principles. For instance,
principles related to secure defaults, fails-safes, access controls, and the
zero trust model~\cite{Ruohonen25JISA} seem to align quite well with the CRA's
new requirements.

Further work is required also on risk analysis and auditing. While the details
are still being worked on in the EU, a good topic related to risk analysis would
be an evaluation of the suitability of existing standards for auditing, whether
internally or externally, a fulfillment of the CRA's essential
requirements. Although there is existing work in this area~\cite{ENISAJRC24},
risk analysis itself seems to have received less attention. As was noted in
Subsection~\ref{subsec: risks}, risk analysis should guide the prioritization of
the essential cyber security requirements and their implementations. As was also
noted, however, a risk analysis requires a case or a uniform domain; universal
guidelines are impossible to postulate because of the CRA's~wide scope.

Auditing would be a good topic also otherwise because cyber security audits are
only rarely conducted according to some industry
surveys~\cite{Rindell21IST}. Regarding the survey results referenced, it seems
that the CRA patches also some other notable limitations in industry
practices. Although costs are certainly a factor, also attack surface analysis
and security reviews more generally seem to be only occasionally done in the
industry~\cite{Rindell21IST}. In light of such observations, another industry
survey about the CRA specifically would also shed light on the potential
challenges ahead regarding~compliance.

As already remarked, it be that the challenges are sectoral, specific to some
particular products, or both. For instance, industrial control systems and
supervisory control and data acquisition (SCADA) systems have been seen to
suffer from many of the issues addressed by the CRA. That is, among other
things, these systems do not always have robust authentication and authorization
methods, remote access contains particular risks, updates are often lacking, and
monitoring and logging are oftentimes insufficient~\cite{Ali24}. These and other
real or perceived challenges correlate with real or perceived problems in
addressing them. For instance: due to the criticality of the systems in many
industry sectors, including those related to countries' critical
infrastructures, integrity guarantees and pre-delivery and post-delivery
verification of updates are likely of significant importance. That is, faulty
updates alone are likely a high-impact risk with these systems and in these
sectors.

\balance
\bibliographystyle{abbrv}

\end{document}